\documentstyle[aps,prb,psfig,floats]{revtex}

\begin{document}

\draft \voffset 1cm

\twocolumn[\hsize\textwidth\columnwidth\hsize\csname %
@twocolumnfalse\endcsname

\title{The normal state scattering rate in high-$T_{c}$ cuprates}
\author{N.E.~Hussey}
\address{H.~H.~Wills Physics Laboratory, University of Bristol, Bristol, BS8 1TL, United Kindgom.}
\date{\today}
\maketitle

\begin{abstract}
I present a new phenomenological model for the normal state transport properties of high-$T_c$ cuprates. In
particular, I identify a form of scattering rate that may account, qualitatively and quantitatively, for the
normal state (magneto)-transport properties of Tl$_2$Ba$_2$CuO$_{6+\delta}$ and
Bi$_2$Sr$_2$CaCu$_2$O$_{8+\delta}$ from optimal doping through to the overdoped side of the phase diagram. The
form of the scattering rate is also consistent with features seen in photoemission spectroscopy in
Bi$_2$Sr$_2$CaCu$_2$O$_{8+\delta}$ and offers a new intuitive way to understand the evolution of the temperature
dependence of the inverse Hall angle with disorder and with carrier concentration.
\end{abstract}

 ]

\section{Introduction}

The search for a consistent description for the anomalous normal state transport properties of high-$T_c$
cuprates (HTC) has been a long and fascinating journey. Many phenomenological models seeking to identify the
appropriate form for the normal state scattering rate have been offered but none have captured the complete
picture. From the very beginning, it was clear that the $T$-linear behavior of the in-plane resistivity
$\rho_{ab}$, extending over an anomalously broad range of temperature for optimally doped (OP) cuprates, would
be difficult to reconcile within a conventional Fermi-liquid (FL) framework.\cite{gurvitch87} Later observations
of the $T$$^2$ dependence and impurity dependence of the inverse Hall angle cot$\Theta_H$,\cite{chien91}
together with the modified Kohler's rule (in-plane orbital magnetoresistance (MR) $\Delta\rho_{ab}/\rho_{ab}
\sim$ tan$^2\Theta_H$),\cite{harris95} seemed to constitute an insurmountable challenge to FL theory.

Several notable attempts to explain the anomalous transport behavior of HTC within a modified FL scenario
followed, each one incorporating a particular form for the (strong) anisotropy of the transport scattering rate
within the plane arising out of some form of coupling to a singular bosonic mode; be it spin
fluctuations,\cite{carrington92,stojkovic96} charge fluctuations\cite{castellani95} or $d$-wave superconducting
fluctuations.\cite{ioffemillis98} In each case, however, inconsistencies with certain aspects of the
experimental data were exposed.

Given these failings, other more exotic models, based on a non-FL ground state, gained prominence within the
community; most notably the two-lifetime picture of Anderson\cite{anderson91} and the Marginal-Fermi-liquid
(MFL) model of Varma and co-workers.\cite{varma89,varma01} The two-lifetime model has been extremely insightful
for understanding the experimental situation in OP cuprates, but has yet to provide an explanation for the
results of Angle-Resolved Photoemission Spectroscopy (ARPES) nor the evolution of the transport phenomena across
the HTC phase diagram. The MFL hypothesis assumes that optimum $T_c$ lies in proximity to a quantum critical
point and consequently, quasi-particles are ill defined everywhere on the Fermi surface with a self-energy that
is governed simply by the temperature scale. In contrast to the two-lifetime picture, this model is consistent
with certain aspects of ARPES measurements,\cite{valla99} though not all. Most notably, the recently reported
evidence of a kink in the energy dispersion\cite{lanzara01} seems at odds with a scenario based on quantum
criticality. Yet again, predictions of the MFL theory have so far been confined to optimum doping.

In this paper, I propose a new model for the normal state scattering rate that offers an alternative route to
understanding the HTC phenomenology. The salient features of the model are presented in section II. The model
has as its basis, two novel but physically reasonable assumptions. Firstly, the dominant scattering rate is
assumed to vary as $T^2$ everywhere on the Fermi surface, but in contrast to previous approaches, both the
elastic $and$ the inelastic parts of the self-energy contain strong (doping dependent) basal plane anisotropy.
The second key assumption is that this anisotropic scattering rate eventually saturates (on different regions of
the Fermi surface at different temperatures) at a value equivalent to the Mott-Ioffe-Regel criterion for
coherent band transport. In section III A, the model is compared with experimental data on
Tl$_{2}$Ba$_{2}$CuO$_{6+\delta}$ (Tl2201) at different doping concentrations. Excellent qualitative and
quantitative agreement is found between the model and experimental data for $\rho_{ab}$($T$), cot$\Theta_H$($T$)
and significantly, the magnitude and $T$-dependence of $\Delta\rho_{ab}/\rho_{ab}$($T$), with relatively few
fitting parameters. The model goes beyond previous proposals in that it can also account for the evolution of
the observed normal state behavior across the overdoped (OD) region of the phase diagram from optimum doping. In
section III B, the model is applied to recent ARPES measurements of the quasi-particle scattering rate and
normal state transport data for Bi$_2$Sr$_2$CaCu$_2$O$_{8+\delta}$ (Bi2212) and comparisons are made with the
single-layer Bi$_2$Sr$_{2-x}$La$_x$CuO$_{6+\delta}$ (Bi2201). In particular, the model is shown to provide a
natural explanation for the non-universal power law behavior of cot$\Theta_H$($T$) observed in the Bi-based
cuprates.\cite{andomurayama99,konstant00} Finally, in section IV, justification arguments for this particular
choice of scattering rate are presented and the implications of this phenomenological model for the physics of
HTC across the phase diagram are discussed, with suggestions for future study. In the process, I challenge two
of the previously established paradigms of the normal state in HTC; namely the linear low-$T$ behavior of
$\rho_{ab}$($T$) and the relevance of the non-saturating $\rho_{ab}$($T$) observed at higher temperatures.

\section{Model}

There is strong experimental evidence from both ARPES\cite{yusof01} and $c$-axis MR\cite{hussey96} that the
dominant scattering mechanism in HTC is highly anisotropic within the basal plane, being strongest at the
so-called `anti-nodal' points at ($\pi$, 0) and weakest at the `nodal' points ($\pi$, $\pi$). In an anisotropic
2D metal, Ong has shown that the Hall angle tan$\Theta_{H}$($T$) will be dominated by those regions of the Fermi
surface where the scattering is weakest.\cite{ong91} If the anisotropy is sufficiently high,
cot$\Theta_{H}$($T$) should therefore reflect the $T$-dependence of the scattering rate at the nodal points.
Given that in the majority of cuprates, cot$\Theta_{H}$($T$) $\sim A + BT$$^{2}$, I take this to represent the
intrinsic $T$-dependence of the dominant scattering rate in HTC. A key stipulation of the present model is that
the scattering rate has this same $T$-dependence $everywhere$ on the Fermi surface, but has four-fold symmetry,
being maximum along ($\pi$, 0).\cite{yusof01,hussey96} Combined with an elastic scattering rate
$\Gamma_{0}$($\phi$), which may also contain 4-fold anisotropy,\cite{varma01} the `ideal' scattering rate
$\Gamma_{\rm{ideal}}$($\phi$) is written as

\[
\Gamma_{\rm{ideal}}(\phi, T) = \Gamma_{0}(\phi) + \Gamma_{T}(\phi, T)
\]
where
\[
\Gamma_{0}(\phi) = \alpha(1 + c\cos^{2}2\phi)
\]
and the $T$-dependent scattering rate
\[
\Gamma_{T}(\phi, T) = \beta(1 + e\cos^{4}2\phi)T^{2}
\]

Here $\phi$ represents the angle between the in-plane Fermi wave vector and ($k_{x}$, 0) whilst $c$ and $e$ are
the anisotropy factors for the impurity and $T$-dependent scattering rates respectively. The coefficients
$\alpha$ and $\beta$ can be roughly estimated from Hall angle measurements. The particular angular dependencies
for $\Gamma_{0}$($\phi$) and $\Gamma_{T}$($\phi$) shown above are chosen to reflect the four-fold symmetry of
$\Gamma_{\rm{ideal}}$($\phi$) and to be consistent with ARPES measurements, as outlined in section III B.
$\Gamma_{\rm{ideal}}$($\phi$) is largest (= $\alpha$(1 + $c$) + $\beta$(1 + $e$)$T$$^2$) along ($\pi$, 0) and
smallest, but still finite (= $\alpha$ + $\beta T$$^2$), along ($\pi$, $\pi$).

The second key feature of the model is the application of the Mott-Ioffe-Regel (MIR) limit to HTC. In its basic
form, the MIR criterion states that the quasi-particle mean free path $\ell$ cannot become smaller than the
lattice spacing $a$, since at this point, the quasi-particles lose their coherence and conventional Boltzmann
transport analysis becomes irrelevant. When $k$-dependence is introduced, different regions of the Fermi surface
become incoherent at different temperatures. Likewise, the scattering rate at different points on the Fermi
surface will saturate (at the MIR limit) at different temperatures.

The first evidence for resistivity saturation in metals, consistent with the MIR limit, was reported in the A-15
superconductors.\cite{fiskwebb76} The $T$-dependence of the resistivity was found to fit extremely well to a
simple `parallel-resistor' model,\cite{wiesmann77} suggesting that the ideal resistivity (i.e. in the absence of
saturation) was somehow shunted by a large saturation resistivity $\rho_{\rm{max}}$ corresponding to $\ell = a$.
Using the MIR criterion as his guide, Gurvitch later argued\cite{gurvitch81} that there must be a minimum time
below which no scattering event can take place, i.e. a minimum distance $a$ over which each carrier gains
additional drift velocity before being scattered, and from this derived an elegant understanding of the
appropriateness of the parallel-resistor model for resistivity saturation.

In the spirit of Gurvitch's picture, I introduce for HTC a maximum scattering rate (minimum lifetime)
$\Gamma_{\rm{max}}$($\phi$), equivalent to $\ell = a$, and define an \emph{effective} scattering rate
$\Gamma_{\rm{eff}}$($\phi$, $T$) at each point on the Fermi surface of the parallel-resistor form, i.e.

\[
\frac{1}{\Gamma_{\rm{eff}}(\phi, T)} = \frac{1}{\Gamma_{\rm{ideal}}(\phi, T)} +
\frac{1}{\Gamma_{\rm{max}}(\phi)}
\]

$\Gamma_{\rm{eff}}$($\phi$, $T$) is then inserted into the Jones-Zener expansion of the standard Boltzmann
transport equation for a quasi-2D Fermi surface to extract all measurable transport coefficients. A derivation
of the appropriate expressions is included as an Appendix. Since the Fermi velocity $v_{F}$ (= 1 - 3 x 10$^{5}$
ms$^{-1}$) has already been determined by ARPES, the value of $\Gamma_{\rm{max}}$ can be fixed directly by the
MIR criterion ($\ell = a$) giving $\Gamma_{\rm{max}}$ in the range 250 - 750 THz (150 - 450 meV). As will be
shown, this derivation of $\Gamma_{\rm{eff}}$($\phi$, $T$) is sufficient to explain the many unusual
$T$-dependencies that appear in the normal state transport in HTC.

\begin{figure}
\centerline{\psfig{figure=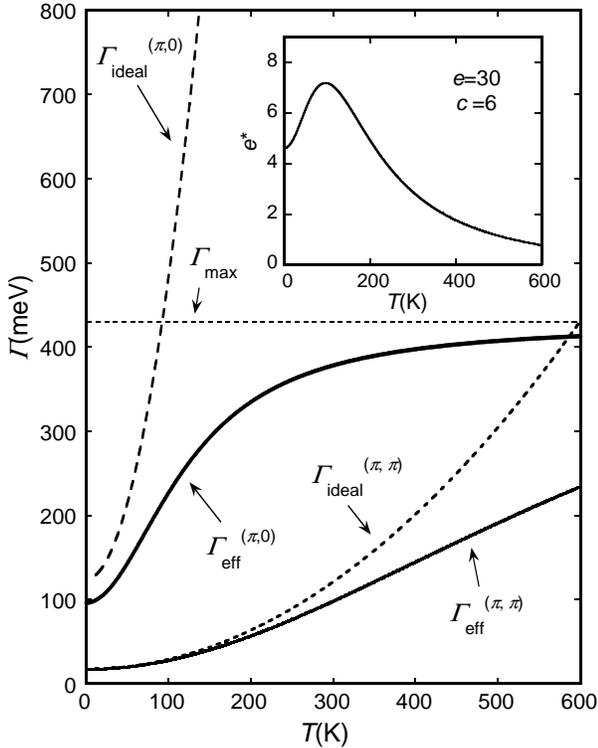,width=8cm}} \vskip 0cm \caption{$T$-dependence of the ideal and
effective scattering rates for ($\pi$, 0) and ($\pi$, $\pi$) within the model. The parameters used are given in
the text. Inset: $T$-dependence of the effective anisotropy $e$* = $\Gamma_{\rm{eff}}$$^{(\pi,
0)}$/$\Gamma_{\rm{eff}}$$^{(\pi, \pi)}$ - 1 for the same parameters} \label{1}
\end{figure}

The $T$-dependencies of $\Gamma_{\rm{eff}}$($\phi$, $T$) for the two momentum directions ($\pi$, 0) and ($\pi$,
$\pi$) are illustrated schematically in Fig.\ \ref{1}. For concreteness, the parameters used in this simulation
are those appropriate for OP Tl2201, namely $\alpha$ = 18 meV, $\beta$ = 1.15 x 10$^{-3}$ meV/K$^{2}$, $c$ = 6,
$e$ = 30 and $\Gamma_{\rm{max}}$$^{(\pi, 0)}$ = 435 meV. Along ($\pi$, 0), $\Gamma_{\rm{ideal}}$($T$) reaches
$\Gamma_{\rm{max}}$ at $T_c$ $\sim$ 90K, at which point, the anti-nodal quasi-particles will become incoherent.
Due to the presence of the `shunt' $\Gamma_{\rm{max}}$, however, $\Gamma_{\rm{eff}}$($T$) is always smaller than
$\Gamma_{\rm{ideal}}$($T$) and approaches saturation much more gradually. For the quasi-particles at ($\pi$,
$\pi$) on the other hand, $\Gamma_{\rm{ideal}}$($T$) reaches $\Gamma_{\rm{max}}$ at a much higher temperature
($T \sim$ 600K). Hence, at the nodal points, well-defined coherent quasi-particles exist at all relevant
temperatures. Note that $\Gamma_{\rm{eff}}$($T$) at ($\pi$, $\pi$) is quasi-linear over a wide temperature
range. The `effective' anisotropy factor $e$* (= $\Gamma_{\rm{eff}}$$^{(\pi, 0)}$/$\Gamma_{\rm{eff}}$$^{(\pi,
\pi)}$ - 1) is significantly lower than $e$ for all finite $T$, reaching a maximum value $\sim$ 7 at the
temperature where $\Gamma_{\rm{ideal}}$$^{(\pi, 0)}$ first crosses $\Gamma_{\rm{max}}$ (see inset to Fig.\
\ref{1}). The overall effect is a transport scattering rate $\Gamma_{\rm{eff}}$($\phi, T$) whose anisotropy
initially grows with increasing $T$ but then gradually becomes more isotropic at higher temperatures as the
scattering rate at different regions of the Fermi surface tends towards saturation. In this way, the
$T$-dependence of $e$*($T$) mimics that of the Hall coefficient $R_H$($T$) in HTC.

\section{Fitting}

\subsection{Application to Tl2201}

\begin{figure}
\centerline{\psfig{figure=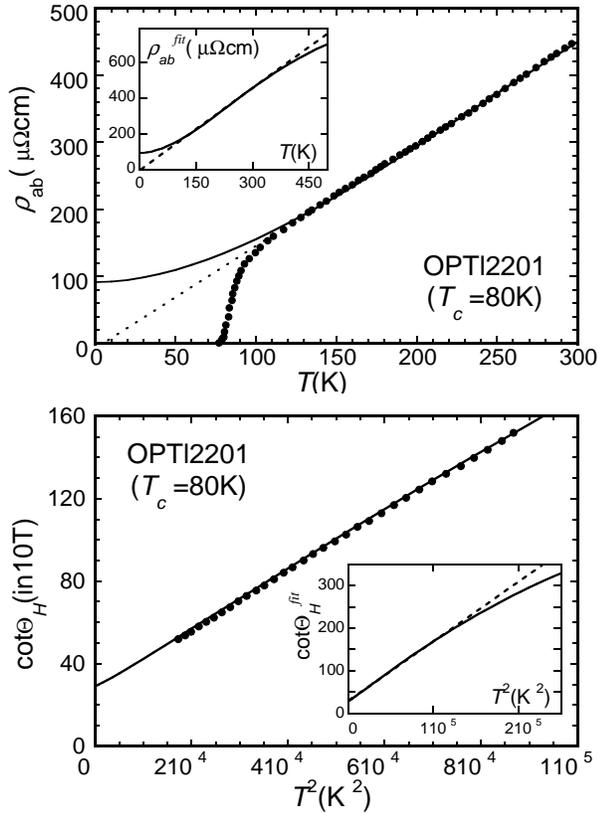,width=8cm}} \vskip 0cm \caption{Top: $\rho_{ab}$($T$) (closed circles)
and $\rho_{ab}^{fit}$($T$) (solid line) for OP Tl2201.\cite{tylermac97} The dotted line is an extrapolation of
$\rho_{ab}$($T$) to $T$ = 0K. Bottom: cot$\Theta_H$($T$) (closed circles) and cot$\Theta_H^{fit}$($T$) for the
same crystal.\cite{tylermac97} Insets: The same fits up to 500K, showing the onset of saturation at higher $T$.}
\label{2}
\end{figure}

Tl2201 is an ideal system for modelling the normal state transport properties in HTC for several important
reasons; (i) Tl2201 is a single band cuprate, and therefore does not contain any secondary conducting subsystem,
such as the CuO chains that complicate Hall and/or MR data in YBa$_2$Cu$_3$O$_{7-\delta}$ (YBCO); (ii) having a
single CuO$_{2}$ plane, bilayer coupling is irrelevant; (iii) the doping range of Tl2201 covers only the OD
region of the phase diagram thereby avoiding any further complications due to the normal state pseudogap; (iv) a
$T$-dependent anisotropy in the in-plane scattering rate has already been observed in OD Tl2201 ($T_c$ = 30K)
from $c$-axis MR measurements;\cite{hussey96} and finally (v) $\rho_{ab}$($T$) and cot$\Theta_{H}$($T$) exhibit
the very clean linear and quadratic $T$-dependencies characteristic of OP cuprates as shown in Fig.\ \ref{2}
(the experimental data are represented by closed circles).\cite{tylermac97,tyler97} One surprising feature of
the experimental data is the small \emph{negative} zero-temperature intercept in $\rho_{ab}$($T$) (extrapolation
indicated by a dotted line) co-existing with a large \emph{positive} intercept in cot$\Theta_H$($T$). Line fits
to the experimental data between 130K and 300K actually give $\rho_{ab}$($T$) = -10 + 1.56$T$ ($\mu\Omega$cm)
and cot$\Theta_H$($T$) = 27 + 0.0014$T$$^2$ (in 10 Tesla).\cite{tylermac97} This subtle but important feature
has often been overlooked but as will be shown, is a natural consequence of the form of the scattering rate
presented here.

To complete the Boltzmann transport analysis, I introduce a Fermi wave vector $k_F$ and Fermi velocity $v_F$
consistent with band structure calculations,\cite{singhpickett92} i.e. $k_F$($\phi$) = 6.5(1 + 0.15
sin$^2$2$\phi$) $\AA^{-1}$ and $v_F$($\phi$) = 2.5(1 + 0.2 sin$^2$2$\phi$) x 10$^5$ ms$^{-1}$. This expression
for $k_F$($\phi$) is consistent with a hole doping concentration $p$ = 0.16. The small anisotropy factors do not
play a significant role in the following analysis and for simplicity, both $k_F$($\phi$) and $v_F$($\phi$) are
assumed to be independent of doping concentration. The corresponding saturation scattering rate $\Gamma_{\rm
max}$($\phi$) = 650(1 + 0.2 sin$^2$2$\phi$) THz or 435(1 + 0.2 sin$^2$2$\phi$) meV (since $a$ = 3.86$\AA$). Note
that the value of $\Gamma_{\rm max}$ is entirely determined by our choice of $v_F$ and contains the same
anisotropy factor as $v_F$ to ensure that the condition $\ell$ = $a$ is satisfied everywhere. Again, these
values are kept constant for all doping concentrations investigated here. Thus in the ensuing analysis, there
are only four adjustable parameters, namely the two anisotropy parameters $c$ and $e$ and the coefficients of
$\Gamma_{\rm{ideal}}$($T$), $\alpha$ and $\beta$. The choice of $\alpha$ and $\beta$ is somewhat restricted
however since according to the Ong representation, $\alpha$ and $\beta$ should be comparable with the
coefficients $A$ and $B$ determined from cot$\Theta_H$($T$). These four parameters are then adjusted to obtain
the best global fits, labelled hereafter $\rho_{ab}^{fit}$($T$), cot$\Theta_H^{fit}$($T$) and
$\Delta\rho_{ab}^{fit}$/$\rho_{ab}^{fit}$($T$), to the three independent transport properties $\rho_{ab}$($T$),
cot$\Theta_H$($T$) and $\Delta\rho_{ab}/\rho_{ab}$($T$). Finally, scaling corrections of the order 30$\%$ were
allowed in the fits to account for any systematic errors in the experimentally determined quantities.

The solid lines in Fig.\ \ref{2} are $\rho_{ab}^{fit}$($T$) and cot$\Theta_H^{fit}$($T$) for OP Tl2201. The
fitting parameters used are $\alpha$ = 27 THz, $\beta$ = 0.00175 THz/K$^2$, $c$ = 6 and $e$ = 30. Above 130K,
both the linear resistivity and the quadratic Hall angle are reproduced with slopes and (extrapolated) $T$ = 0
intercepts in excellent agreement with the experimental data. Below 130K, $\rho_{ab}^{fit}$($T$) deviates from
linearity, approaching $T$$^2$ as $T$ $\rightarrow$ 0, whilst $\rho_{ab}$($T$) actually shows a downward
deviation from linearity below 130K, presumably due to the onset of superconducting fluctuations which always
dominate the intrinsic behavior of the normal state $\rho_{ab}$($T$) near $T_c$. Note that the $T$ dependence of
$\Gamma_{\rm{ideal}}$($T$) (= 27 + 0.00175$T$$^2$) is similar to that of cot$\Theta_{H}$($T$) (= 27 +
0.0014$T$$^2$), in good agreement with Ong's picture.\cite{ong91}

The insets in Fig.\ \ref{2} show $\rho_{ab}^{fit}$($T$) and cot$\Theta_H^{fit}$($T$) over an extended
temperature range up to 500K. The influence of $\Gamma_{\rm{max}}$ on the high-$T$ behavior is manifest as a
slight deviation from linearity above 360K, indicating the onset of saturation in $\Gamma_{\rm{eff}}$($T$). The
fact that this tendency towards saturation is not actually observed in $\rho_{ab}$($T$) indicates that the model
does not capture all the essential physics. However, as I argue in section IV, there are several possible
factors that might lead to a non-saturating resistivity in HTC which go beyond conventional Boltzmann transport
analysis and may mask the intrinsic behavior of the scattering rate at higher temperatures.

\begin{figure}
\centerline{\psfig{figure=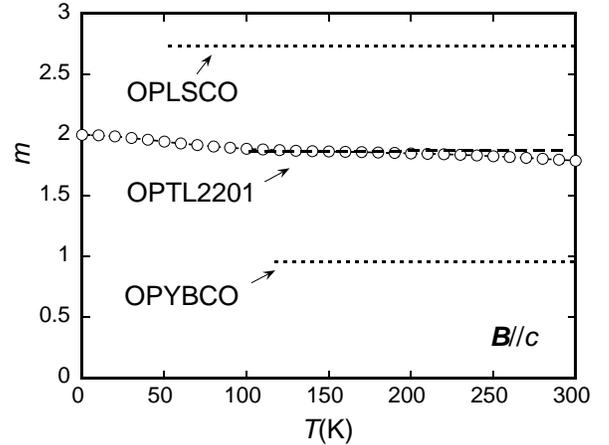,width=8cm}} \vskip 0cm \caption{\emph{Dashed line}: Modified Kohler's
plot $m$($T$) for OP Tl2201 (where $m$$^{2}$ =
($\Delta\rho_{ab}$/$\rho_{ab}$)/tan$^{2}\Theta_{H}$).\cite{tyler97} The open circles represent $m^{fit}$($T$)
obtained with the same fitting parameters as used in Fig. \ \ref{2}. \emph{Dotted lines}: $m$($T$) for OP LSCO
and YBCO.\cite{harris95}} \label{3}
\end{figure}

The modified Kohler's rule $\Delta\rho_{ab}$/$\rho_{ab}$ = $m$$^{2}$ tan$^2\Theta_H$ is also obeyed in OP
Tl2201, with $m \sim$ 1.9 down to 130K,\cite{tyler97} as shown by the dashed line in Fig.\ \ref{3}. (Below this
temperature, superconducting fluctuations again begin to dominate.) This value lies in between the
experimentally determined values for OP La$_{2-x}$Sr$_x$CuO$_4$ (LSCO)\cite{harris95} and OP YBCO\cite{harris95}
indicated by dotted lines in Fig.\ \ref{3}. The open circles in the figure are $m^{fit}$($T$) for OP Tl2201
calculated using the same model parameters that were used in Fig.\ \ref{2}. As can be seen, $m^{fit}$($T$) has
the correct magnitude and $T$-dependence, decreasing by only 10$\%$ between 0K and 300K. Note that the present
model does not include vertex corrections nor paraconductivity contributions that might influence the form of
$\Delta\rho_{ab}$/$\rho_{ab}$($T$) at low $T$.

The behavior of $m^{fit}$($T$) is determined almost exclusively by the anisotropy parameters $c$ and $e$. (If
$c$ = 0, for example, $\Delta\rho_{ab}^{fit}$/$\rho_{ab}^{fit}$($T$) would fall steeply as $T \rightarrow$ 0.)
The large value of $e$ is surprising, but is necessary, not only to give the correct magnitude for $m$ at finite
temperatures, but also to maintain a robust $T$$^2$ dependence of cot$\Theta_H^{fit}$($T$). Hence, within this
model, the observation of a quadratic Hall angle and the form of $\Delta\rho_{ab}^{fit}$/$\rho_{ab}^{fit}$($T$)
are all signatures of an extremely anisotropic basal plane scattering rate that likewise varies as $T$$^2$. The
$T$-linear resistivity, on the other hand, is simply an artefact of the broad crossover between the
(approximately) $T$$^2$ resistivity at low $T$ and the saturation of $\Gamma_{\rm{eff}}$ at high $T$.

The ability of the present model to reproduce the correct magnitude and $T$-dependence of the orbital MR is to
be contrasted with the `Cold Spots' model \cite{ioffemillis98} where the parameters required to separate the
transport and Hall lifetimes yield an orbital MR that is orders of magnitude too large and has a much stronger
$T$-dependence than shown in Fig.\ \ref{3}. This difficulty is overcome in the present model by the introduction
of the shunt $\Gamma_{\rm{max}}$ that makes the effective anisotropy around the Fermi surface much smaller (see
inset to Fig.\ \ref{1}), thereby reducing both the magnitude and the $T$-dependence of
$\Delta\rho_{ab}^{fit}$/$\rho_{ab}^{fit}$($T$).

\begin{figure}
\centerline{\psfig{figure=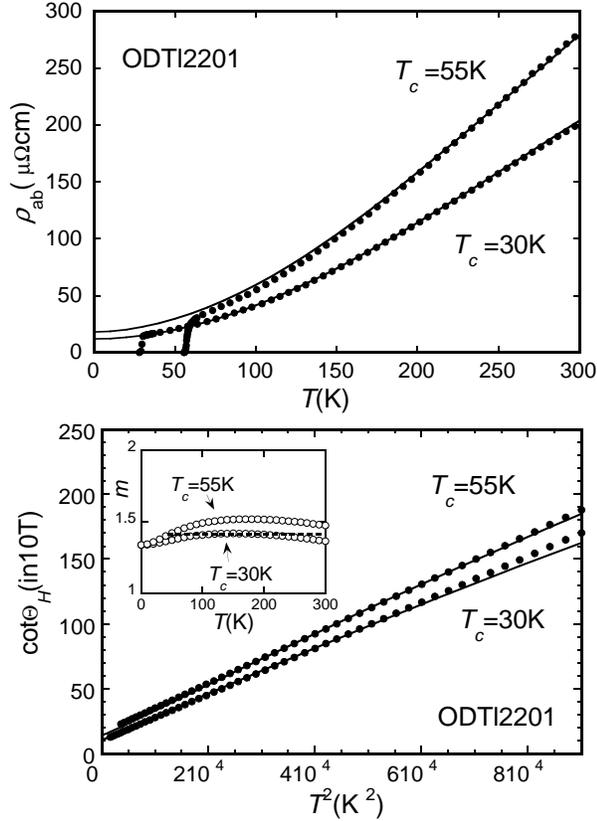,width=8cm}} \vskip 0cm \caption{Top: $\rho_{ab}$($T$)\cite{tyler97}
(closed circles) and $\rho_{ab}^{fit}$($T$) (solid lines) for OD Tl2201 ($T_c$ = 55K and $T_c$ = 30K). Bottom:
cot$\Theta_H$($T$)\cite{tyler97} (closed circles) and cot$\Theta_H^{fit}$($T$) (solid lines) for the same
samples with the same fitting parameters. Inset: $m^{fit}$($T$) (open circles) for the two sets of parameters.
The dashed line represents the experimental data for the 30K crystal.\cite{tyler97}} \label{4}
\end{figure}

The evolution of the normal state transport properties in Tl2201 with doping are summarized in Fig.\ \ref{4}.
The top panel in Fig.\ \ref{4} shows $\rho_{ab}$($T$) (closed circles) and $\rho_{ab}^{fit}$($T$) (solid lines)
for two OD Tl2201 single crystals with $T_c$ values of 55K and 30K.\cite{tyler97} The parameters used in these
fits are $\alpha$ = 9 THz, $\beta$ = 0.0016$T$$^2$ THz/K$^2$, $c$ = 2 and $e$ = 11.5 for the 55K crystal, and
$\alpha$ = 7 THz, $\beta$ = 0.0015$T$$^2$ THz/K$^2$, $c$ = 2 and $e$ = 8 for the 30K crystal. Note that the
anisotropy factors are markedly reduced compared to OP Tl2201, reflecting the increased curvature observed in
$\rho_{ab}$($T$). The bottom panel in Fig.\ \ref{4} shows the corresponding cot$\Theta_H$($T$) and
cot$\Theta_H^{fit}$($T$) for the two samples. In contrast to OP Tl2201, cot$\Theta_H^{fit}$($T$) deviates
slightly from a simple $T$$^{2}$ dependence. In fact, the $T$-dependence is closer to $T$$^{1.95}$ and
$T$$^{1.9}$ for the 55K and 30K crystals respectively. This deviation from a $T$$^{2}$ dependence reflects the
fact that in less anisotropic cuprates, other regions of the Fermi surface are beginning to contribute to the
Hall angle. The inset in the bottom panel of Fig.\ \ref{4} gives the modified Kohler's plot derived for the two
doping levels. As we expect, $m^{fit}$ is lower in these examples compared with OP Tl2201, with values
approaching 1.55 and 1.4 for the 55K and 30K crystal respectively. Experimentally, $m$ is found to be 1.4 for
30K Tl2201 (indicated by a dashed line),\cite{tyler97} again in excellent agreement with the model.

\subsection{Application to Bi2212}

With its improved energy and momentum resolution, ARPES has become a powerful probe of the single-particle
self-energy $\Sigma$ of HTC. The vast majority of work to date has been carried out on the bilayer cuprate
Bi2212, due to its ease of cleaving. One of the most striking features of the emerging spectral function in
Bi2212 is the broad featureless normal state spectrum at ($\pi$, 0) whose lineshape sharpens dramatically into
the well-known peak-dip-hump structure upon cooling below $T_c$. Although interpretation of ARPES spectra is
still a controversial topic,\cite{borisenko01} it has been argued\cite{ioffemillis98,norman98} that this
dramatic rearrangement of the ARPES lineshape indicates the presence of ill-defined quasi-particle states in the
normal state at ($\pi$, 0), rendered incoherent by some form of catastrophic low-energy scattering mechanism.
ARPES data also reveal that the broadening of the quasi-particle self-energy is much less dramatic for
quasi-particles at ($\pi$, $\pi$) and that this basal plane anisotropy is much larger for OP cuprates than for
OD cuprates.\cite{yusof01} All these features indicate a highly anisotropic normal state scattering rate whose
anisotropy grows weaker with overdoping, consistent with the picture described above for Tl2201.

\begin{figure}
\centerline{\psfig{figure=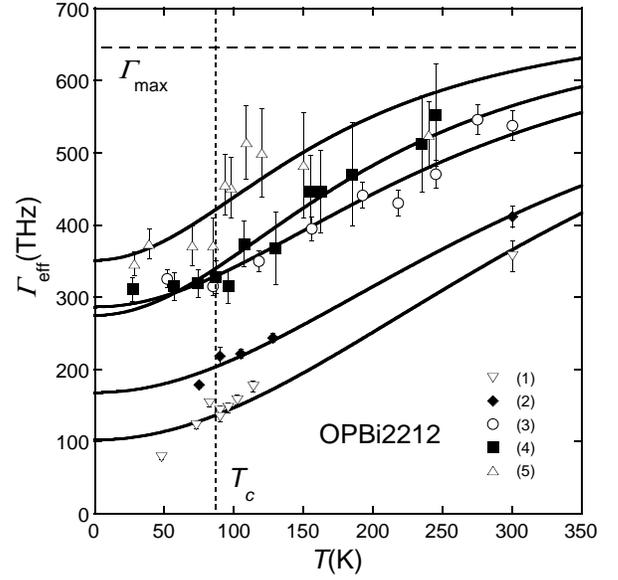,width=8cm}} \vskip 0cm \caption{Quasi-particle scattering rate for OP
Bi2212 extracted from ARPES measurements.\cite{valla00} The various numbers and symbols refer to different
locations on the Fermi surface as described in the text. The solid lines are fits to the individual scattering
rates using the expression for $\Gamma_{\rm{eff}}$($\phi, T$).} \label{5}
\end{figure}

Fig.\ \ref{5} shows the variation of the quasi-particle scattering rate with angle and temperature for an OP
Bi2212 single crystal, extracted from the ARPES data of Valla \emph{et al.}.\cite{valla00} The various symbols
shown in Fig.\ \ref{5} are the same as those used in the original Valla plot of the momentum widths $\Delta k$,
obtained in turn by fitting different momentum distribution curves (MDC) with Lorentzian line shapes. The
quasi-particle scattering rates 1/$\tau$ are obtained using the usual expression 1/$\tau$ =
$\mid$2Im$\Sigma$$\mid$ = $\hbar \Delta kv_k$ where Im$\Sigma$ is the imaginary part of the single-particle self
energy and $v_k$ is the unrenormalized band velocity $\sim$ 4.0 x 10$^5$ ms$^{-1}$. The numbers refer to
locations on the Fermi surface at angles, (1) = 45$^{\circ}$, (2) = 37.5$^{\circ}$, (3) = 25.5$^{\circ}$, (4) =
17.5$^{\circ}$ and (5) = 7$^{\circ}$ away from ($\pi$, 0).

The normal state Fermi velocity, determined at each location from the slope of the dispersion relation near the
Fermi level, can be approximated by the expression $v_F$($\phi$) = 2.5(1 + 0.2 sin$^2$2$\phi$) x 10$^5$
ms$^{-1}$.\cite{valla00} Note that this is the same $v_F$($\phi$) used in the Tl2201 analysis. Similarly,
$\Gamma_{\rm max}$($\phi$) = 650(1 + 0.2 sin$^2$2$\phi$) THz or 435(1 + 0.2 sin$^2$2$\phi$) meV.

\begin{figure}
\centerline{\psfig{figure=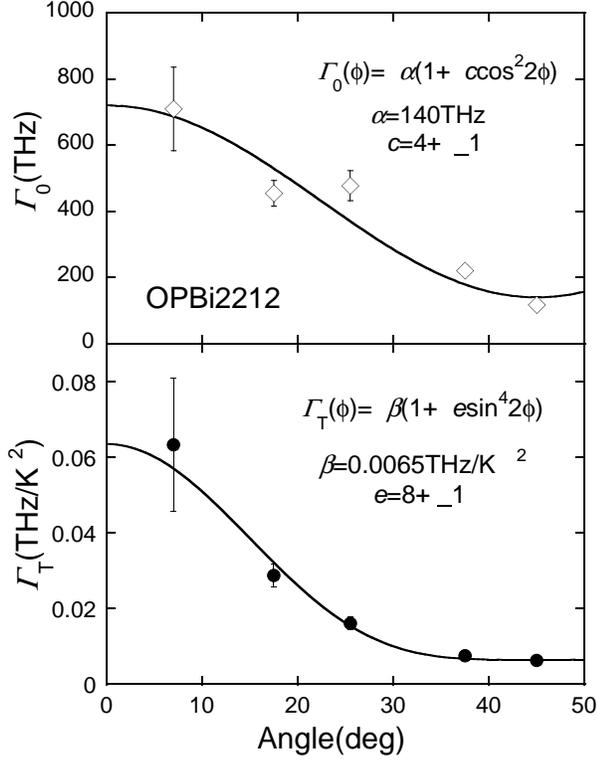,width=8cm}} \vskip 0cm \caption{Angular dependence of $\Gamma_0$ (top
panel) and $\Gamma_T$ (bottom panel) for OP Bi2212 determined from the fits to the MDC curves shown in Fig.\
\ref{5}. The solid lines are best fits whose expressions are shown in the appropriate figures.} \label{6}
\end{figure}

The solid lines in Fig.\ \ref{5} are the fits to the ARPES data using the model scattering rate
$\Gamma_{\rm{eff}}$($\phi, T$). The fits are all reasonable and give angular dependencies for the resulting
$\Gamma_0$($\phi$) and $\Gamma_T$($\phi$) as shown in Fig.\ \ref{6}. The solid lines in Fig.\ \ref{6} are
weighted fits to the expressions $\Gamma_0$($\phi$) = $\alpha$(1 + $c$cos$^2$(2$\phi$) and $\Gamma_T$($\phi$) =
$\beta$(1 + $e$cos$^4$(2$\phi$) with $\alpha$ = 140 THz, $\beta$ = 0.0065 THz/K$^2$, $c$ = 4 $\pm$ 1 and $e$ = 8
$\pm$ 1. Note that $\Gamma_T$($\phi$) is more strongly peaked at ($\pi$, 0) than $\Gamma_0$($\phi$). The reason
these anisotropy factors seem large is the presence of $\Gamma_{\rm max}$. As $\Gamma_{\rm{eff}}$($\phi$, $T$)
approaches $\Gamma_{\rm max}$, $\Gamma_{\rm{ideal}}$($\phi$, $T$) must change significantly in order to create
even a small change in $\Gamma_{\rm{eff}}$($\phi$, $T$). Hence within this model, the `ideal' scattering rate is
actually much more anisotropic than the raw data suggest.

\begin{figure}
\centerline{\psfig{figure=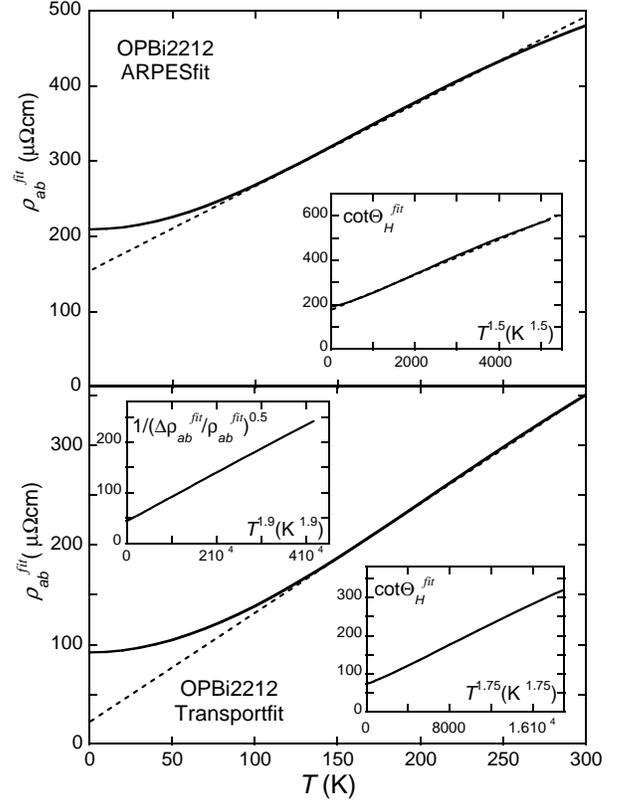,width=8cm}} \vskip 0cm \caption{Top panel: $\rho_{ab}$($T$) and
cot$\Theta_{H}$($T$) (inset) for OP Bi2212 using the fitting parameters derived from the ARPES data. \emph{Solid
circles}: the results from the model, \emph{dashed lines}: guides to the eye. Bottom panel: Best fits to
$\rho_{ab}$($T$) (main figure) cot$\Theta_{H}$($T$) (top inset) and
1/($\Delta\rho_{ab}$/$\rho_{ab}$)$^{0.5}$($T$) (bottom inset) for OP Bi2212. Note the different $T$-dependencies
of the various plots.} \label{7}
\end{figure}

The form of $\Gamma_{\rm{eff}}$($\phi, T$) obtained above is inserted into the Boltzmann equation to obtain a
crude estimate of $\rho_{ab}^{fit}$($T$) and cot$\Theta_H^{fit}$($T$) as shown in the top panel of Fig.\
\ref{7}. The resistivity has a very large zero-temperature offset though is approximately linear from $T_c$ up
to $T$ = 250 K at which point, $\rho_{ab}^{fit}$($T$) tends towards saturation as the effect of $\Gamma_{\rm
max}$ becomes prominent. cot$\Theta_{H}^{fit}$($T$), on the other hand, is found to vary roughly as $A$ +
$BT^{1.5}$. Experimentally, $\rho_{ab}$($T$) in OP Bi2212 shows the ubiquitous $T$-linear behavior, but has a
\emph{positive} zero-temperature intercept, indicative of increased inhomogeneity in this
compound,\cite{watanabe97} whilst cot$\Theta_{H}$($T$) = $A$ + $BT^n$ where $n$ = 1.75 $\pm$
0.05,\cite{beschoten94,konstant00} in contrast to the usual result for OP cuprates, $n$ = 2. Similar behavior is
also observed in Bi2201.\cite{andomurayama99,konstant00}

Whilst the ARPES-derived fits are in reasonable agreement with the experimental data, they are not ideal. In any
case, one does not expect the transport scattering rate to be equivalent to the single-particle scattering rate
obtained by ARPES, since the former should be dominated by large-angle scattering events whilst the latter will
not. The best actual fits to the transport data for OP Bi2212\cite{bi2212data} are shown in the bottom panel of
Fig.\ \ref{7} for the fitting parameters $\alpha$ = 50 THz, $\beta$ = 0.00325 THz/K$^2$, $c$ = 4 and $e$ = 8.
These values give the expected linear resistivity between $T_c$ and 300K with the correct slope and intercept
and cot$\Theta_H^{fit}$($T$) = $A$ + $BT$$^{1.75}$ \emph{over the entire temperature range}, again with the
correct slope and intercept.  The deviation of cot$\Theta_H$($T$) from its simple $T$$^2$ dependence arises from
the increased disorder in Bi2212 that must somehow smear out the anisotropy of the scattering rate around the
Fermi surface. Note that the anisotropy factors are exactly the same as those obtained from the ARPES data,
while $\beta$ in the transport fit is half the ARPES value. The much larger residual scattering rate inferred
from the ARPES spectra is a common feature of photoemission data\cite{valla99} so is not unexpected.
Incidentally, using the same parameters, $\Delta\rho_{ab}^{fit}$/$\rho_{ab}^{fit}$($T$) is found to vary as ($C$
+ $DT$$^{1.9}$)$^{-2}$, also shown in Fig.\ \ref{7}. This is only a prediction from the model at present since
$\Delta\rho_{ab}$/$\rho_{ab}$($T$) has yet to be established experimentally in OP Bi2212.

It should be stressed finally that the ARPES measurements of Valla \emph{et al.}\cite{valla00} are often cited
as strong evidence for a \emph{linear} quasi-particle scattering rate in HTC that is itself responsible for the
$T$-linear resistivity observed at optimal doping. However, as noted by Valla and co-workers, near the
anti-nodal point (label number (5) in Fig.\ \ref{5}), the scattering rate becomes almost $T$-independent. This
tendency towards saturation is an essential part of the present model, and combined with a single scattering
rate that is \emph{quadratic} in temperature everywhere on the Fermi surface, it is able to account for the
quasi-linear scattering rate at the nodes (Fig.\ \ref{5}), the $T$-linear resistivity, the tendency towards
saturation at ($\pi$, 0) and an inverse Hall angle that varies as $A$ + $BT$$^{1.75}$, without requiring any
additional assumptions. Note that in the Cold-Spots model, the scattering rate at the nodes is always quadratic,
whilst in the MFL phenomenology, there is no saturation at ($\pi$, 0).

\section{Discussion}

It is clear that the physical mechanism behind the normal state scattering rate, and \emph{ipso facto}
high-temperature superconductivity, can only be determined once a coherent description of all the normal state
transport properties in HTC is found. In this paper, I have outlined a new phenomenology that shows promising
signs of achieving this goal. However, whilst many of the anomalous features can be explained within this simple
model, the current proposal appears, at first sight at least, to be incompatible with two key aspects of the
experimental situation in HTC; namely the linear resistivity at low $T$ and the absence of resistivity
saturation at high $T$. Therefore, before going on to discuss the possible implications of the model, it is
perhaps worth outlining to the reader how these two major discrepancies might possibly be reconciled.

It has often been claimed that $\rho_{ab}$($T$) in OP cuprates is linear at all finite temperatures. This
long-held view is based largely on the observation back in 1990 of a linear $\rho_{ab}$($T$) in single crystal
Bi2201 extending from 700K down to $T_c$ = 10K.\cite{martin90} If this were indeed the intrinsic behavior, it
would certainly require an explanation outside the realms of conventional FL physics. Important new data on the
latest generation of Bi2201 crystals (with significantly improved $T_c$ values) tell a rather different story
however. By suppressing superconductivity in OP Bi2201 ($T_c$ = 32K) in a 60 Tesla pulsed magnetic field, Ono
$et$ $al.$\cite{ono00} recently found that the linear $\rho_{ab}$($T$) seen at high $T$ (and giving an apparent
zero intercept at $T$ = 0K) crosses over to a higher power $T$-dependence below around 60K, before saturating at
some finite $positive$ $\rho$$_{0}$. Similar behavior has also been reported in OP LSCO.\cite{boebinger96} The
extension of the linear $\rho_{ab}$($T$) to low temperatures in Bi2201 and LSCO (in zero-field) is not therefore
intrinsic and is presumably caused by the onset of fluctuation conductivity, as is evident in the
$\rho_{ab}$($T$) data of Tl2201 and Bi2212. In addition, the (extrapolated) negative intercept in
$\rho_{ab}$($T$) found in the best quality Tl2201 and YBCO crystals\cite{tylermac97,yoshida99} is unphysical and
dictates that $\rho_{ab}$($T$) \emph{must} cross over to a higher power of $T$ at lower temperatures. Such
behavior is incompatible with the MFL hypothesis, for example, but is a natural consequence of the form of
$\Gamma_{\rm{eff}}$($\phi$, $T$) presented here. Perhaps in the light of these new results, the long-standing
assertion that $\rho_{ab}$($T$) obeys a simple linear $T$-dependence at low $T$ should now be re-examined. It
would certainly be interesting to follow the $T$-dependence of OP cuprates with higher $T_c$ to lower
temperatures, as and when larger magnetic fields become available.

I now turn to address the second point, namely the absence of resistivity saturation at higher temperatures. The
MIR limit and the parallel-resistor model have been applied successfully to a wide range of materials over the
years. Their relevance to strongly correlated electron systems such as HTC, however, has been challenged in
recent times.\cite{emerykivelson95} In these so-called `bad metals', $\rho$($T$) does not saturate. Instead
$\rho$($T$) grows approximately linearly with $T$ at high temperatures (typically $T$ $>$ 300K) reaching values
of the order 1 - 10 m$\Omega$cm that correspond to $\ell \ll a$. It is claimed that this absence of resistivity
saturation is a signature of a non-FL ground state at zero temperature.\cite{emerykivelson95} At first sight,
these arguments appear rather compelling. However, the origin of the non-saturating resistivity in these systems
is still poorly understood. More importantly, similar $\rho$($T$) behavior has also been reported in
Sr$_{2}$RuO$_{4}$,\cite{tyler98} a close structural analog of the cuprates which exhibits quantum
oscillations\cite{mackenzie96a} and so forms a well-defined FL ground state at low $T$. The association of
non-saturating $\rho$($T$) with a non-FL ground state is therefore misleading and that some other physical
origin must be sought for the high-$T$ behavior.

Possible sources for the excess resistivity include thermal expansion effects (known to play a significant role
in the organic conductors, for example), a thermally-induced reduction in the density of states or carrier
concentration, and the onset of a `thermal' diffusive (non-Boltzmann) transport regime at high $T$ that sets in
once the concept of a scattering rate is no longer applicable and temperature becomes the only relevant energy
scale in the problem. Determining the origin of this non-saturating resistivity, though desirable, is not
essential for justifying the model however, since the current model is only strictly applicable in the low-$T$
($T$ $<$ 300K) regime, where the majority of the carriers are still well-defined quasi-particles and Boltzmann
transport analysis remains valid.

Any observation of a threshold in the scattering rate would obviously lend significant weight to the arguments
presented here however. In addition to the work of Valla \emph{et al.},\cite{valla00} Norman \emph{et al}. have
recently extracted Im$\Sigma$ along ($\pi$, 0) for slightly OD Bi2212\cite{norman98} and found that it too
saturates above $T_c$ at a value 2Im$\Sigma_{\rm{max}} \sim$ 180 meV, well within the range of validity given by
$\ell = a$. Optical measurements can also be used in principle to separate $\Gamma$ from the conductivity,
though here the situation is not as clear. In particular, $\Gamma$ is found to saturate with increasing
temperature, but NOT with increasing frequency.\cite{puchkov96} Finally, recent analysis of the plasmon spectra
in K$_{3}$C$_{60}$,\cite{goldini01} another system with non-saturating resistivity, has revealed that the
quasi-particle scattering rate itself saturates around $T$ = 500K at a value $\Gamma_{\rm{max}} \sim$ 450 meV.
Whilst more experimental work is clearly required to clarify this effect, these observations do point to the
possibility that the MIR limit may still be a preserve of $all$ metallic systems, in the strictest sense that
$\Gamma$ never exceeds the maximum value $\Gamma_{\rm{max}}$ corresponding to $\ell = a$, and that the excess
resistivity seen in these strongly correlated systems should not be regarded simply as a signature of an
ever-increasing scattering rate. Moreover, since the phenomenology described in this paper has been shown to
provide a coherent description of many of the anomalous normal state transport properties in HTC, further
efforts to separate $\Gamma$($T$) from $\rho$($T$) are strongly recommended.

\begin{figure}
\centerline{\psfig{figure=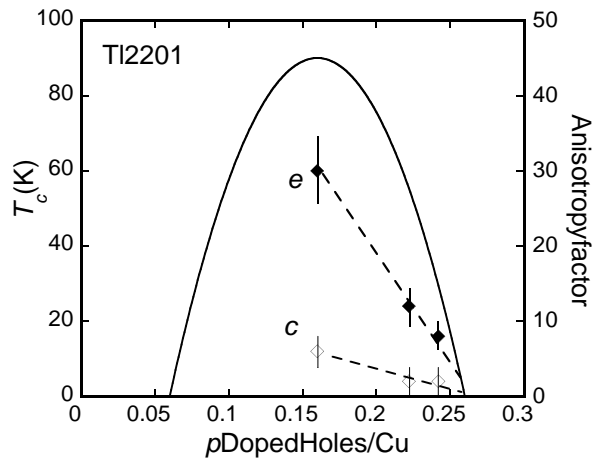,width=8cm}} \vskip 0cm \caption{Variation of the anisotropy factors
$c$ (open diamonds) and $e$ (closed diamonds) with doping concentration $p$. The dashed lines are guides to the
eye. The solid line represents the $T_c$($p$) parabola for HTC.\cite{tallon95} The error bars reflect the range
of fitting parameters that provide a reasonable fit to all three transport coefficients.} \label{8}
\end{figure}

Turning now to the implications of the model, the most striking features of the analysis on Tl2201 are clearly
the large values for $c$ and $e$ and their strong variation with doping, as summarized in Fig.\ \ref{8}. (For
each crystal, the doping level $p$ is obtained from the well-known HTC parabola law
$T_c$($p$)/$T_{c}^{\rm{max}}$ = 1 - 82($p$ - 0.16)$^{2}$,\cite{tallon95} shown as a solid line in Fig.\
\ref{8}.)

The large value of $c$, particularly at optimum doping, is somewhat surprising, though it is not inconsistent
with the picture emerging from ARPES measurements on Bi2212.\cite{valla00} Recently it has been
argued\cite{varma01} that scattering on impurities located \emph{off} the CuO$_{2}$ planes may be the source of
this four-fold anisotropy in $\Gamma$$_{0}$($\phi$). In the case of tetragonal Tl2201, such interplane
randomness might be induced by Tl/Cu site substitutions as well as by interstitial oxygen. It is not clear at
this stage why such large anisotropy would also appear in the transport scattering rate, which typically depends
only on large-momentum transfers. However, as noted by Varma and Abrahams,\cite{varma01} such anisotropic
elastic scattering may well play a key role in the $magneto$-transport behavior. Indeed, the large value of $c$
is only required in the fitting here to maintain the constancy of $m^{fit}$($T$) down to low temperatures.

The $T$-dependent anisotropy factor $e$ is found to drop steeply from optimal doping, scaling approximately with
$T_c$ and approaching zero close to the doping level where superconductivity vanishes. This is to be expected
perhaps, since the separation of transport and Hall lifetimes is observed to vanish in the most highly OD
samples.\cite{note1} The correlation between $e$($p$) and $T_c$($p$) on the OD side is intriguing and suggests
that the mechanism giving rise to this strongly anisotropic scattering rate might also be associated with the
anisotropy ($d$-wave pairing symmetry) of the superconducting order parameter.

As illustrated in Fig.\ \ref{1}, $\Gamma_{\rm{ideal}}$$^{(\pi, 0)}$($T$) for OP Tl2201 is found to cross the MIR
limit close to $T$ = $T_c$. ($\Gamma_{\rm{eff}}$$^{(\pi, 0)}$($T$) will obviously be lower than the MIR limit
due to the presence of the shunt $\Gamma_{\rm{max}}$.) The same is also true for OP Bi2212. The picture that
emerges therefore is one in which the quasi-particles at ($\pi$, 0) are approaching an incoherent state at $T$ =
$T_c$, consistent with what is observed in ARPES, and this may well have implications for the onset temperature
of superconductivity on the underdoped (UD) side of the phase diagram. More explicitly, if $e$ continues to grow
with decreasing $p$, the temperature at which the anti-nodal quasi-particles become incoherent will decrease
accordingly. What ultimately causes the suppression of $T_c$ on the UD side of the phase diagram is still an
open issue, and I cannot address that important question here, since clearly, the physics of the UD cuprates is
also heavily influenced by the opening of the pseudogap which cannot yet be accounted for within this simple
phenomenological model. Nevertheless, the onset of incoherent quasi-particles at $T$ = $T_c^{\rm{max}}$ is an
intriguing observation from the above analysis, hints at a new interpretation of the $T_c$($p$) parabola across
the HTC phase diagram and should be investigated further.

\begin{figure}
\centerline{\psfig{figure=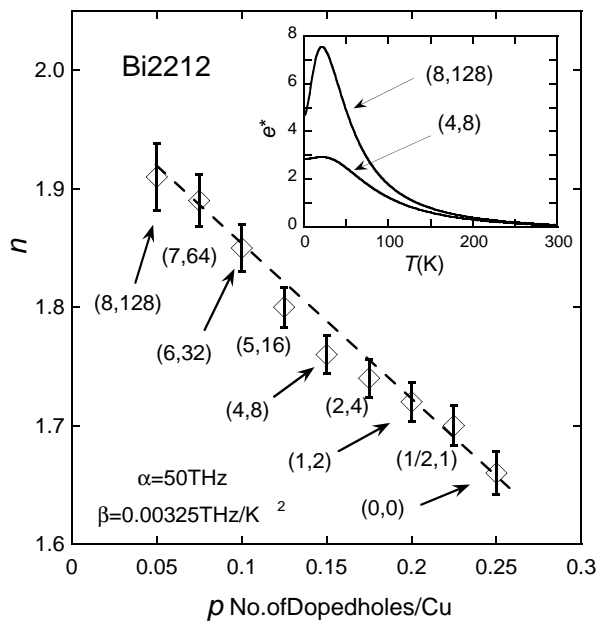,width=8cm}} \vskip 0cm \caption{Simulation of the variation of
$n$($p$) for Bi2212 where $n$ is the coefficient of inverse Hall angle cot$\Theta_H$($T$) = $A$ + $BT^n$. The
anisotropy factors ($c$, $e$) used in the fitting routines are indicated in parentheses. The dashed line is a
guide to the eye.\cite{konstant00} Inset: Effective anisotropy $e$*($T$) for two selected points.} \label{9}
\end{figure}

Some possible insight into the nature of the scattering on the UD side can be gained from further analysis of
the Hall angle in Bi2212. In recent detailed transport studies of the Bi-based cuprates Bi2212 and Bi2201,
cot$\Theta_H$($T$) was found to vary as $T^n$ with $n$ decreasing from $n \sim$ 2 to $n \sim$ 1.6 - 1.7 as one
moves from the UD to the OD regime. This non-universal power law behavior in cot$\Theta_H$($T$) reveals a new
layer of complexity in the phenomenology of normal state transport in HTC that has yet to be addressed. It was
noted earlier for Tl2201, that $n$ decreases on the OD side as $c$ and $e$ are reduced. Given that at optimal
doping, $n$ $\sim$ 1.8 lies roughly in the middle of the variable range for Bi2212, it is tempting to attribute
this variation in $n$($p$) purely to changes in $c$ and $e$ across the phase diagram. A simple example of how
this might come about is illustrated in Fig.\ \ref{9} where $c$ and $e$ have been varied systematically to
simulate the experimentally observed variation of $n$ with doping,\cite{konstant00} whilst keeping $\alpha$ and
$\beta$ fixed at the values extracted from the transport fits for OP Bi2212 in Fig.\ \ref{7}. Intriguingly, in
the isotropic case $c$ = $e$ = 0, $n$ $\sim$ 1.65, in agreement with the experimental data. As $c$ and $e$ are
increased, $n$ rises as expected, but only approaches $n$ = 2 for extremely large values of $e$. Whilst this
example is not completely rigorous, it does suggest that with more careful analysis, the systematic variation of
$n$ with doping in Bi2212 and Bi2201, and its range of values, might be understood within this simple model. It
also implies that the anisotropy of the $T$$^2$ scattering rate must increase (almost exponentially) with
decreasing doping on the UD side, reaching values greater than 100 for $p \sim$ 0.05! Indeed, within the present
model, it is the \emph{only} way to account for the observation of cot$\Theta_H$($T$) = $A$ + $BT$$^2$ in UD
Bi2212. However, as illustrated in the inset of Fig.\ \ref{9}, the effective anisotropy $e$*($T$) above 100K is
almost indistinguishable for the OP and the most UD cases. Hence, whilst $e$ may increase dramatically, the more
experimentally accessible quantity $e$* does not.

Comparison of the fitting parameters for OP Bi2212 and OP Tl2201 reveal another important aspect of the
modelling. The degree of anisotropy in OP Bi2212 ($c$ = 4, $e$ = 8) is markedly reduced compared with that in OP
Tl2201 ($c$ = 6, $e$ = 30) whilst the coefficients of $\Gamma_{\rm{ideal}}$($T$) ($\alpha$ = 50 (27) and $\beta$
= 0.00325 (0.00175) for OP Bi2212 (OP Tl2201)) are increased. This implies that the strength of the scattering
rate, averaged around the Fermi surface, is very similar in the two cases, but the higher degree of disorder and
inhomogeneity in Bi2212, evidenced by the larger value of $\alpha$ (the level of impurity scattering at the
nodal points) acts to smear out this strong anisotropy.

Whilst I do not wish to speculate at this stage on the origin of $\Gamma_{\rm{eff}}$($\phi$, $T$), there are
still some important remarks that can be made from the above analysis. Firstly, the strong anisotropy and
$T$$^2$ dependence of $\Gamma_{\rm{eff}}$($\phi$, $T$) strongly suggests that the dominant scattering rate in
HTC is electronic in origin and that electron-phonon scattering plays only a minor role in the normal state
transport properties of HTC. The large increase in the quasi-particle lifetime below $T_c$ observed in
conductivity measurements\cite{yu92,bonn93} is consistent with the disappearance of this catastrophic electronic
scattering mechanism below $T_c$. Secondly, the large value for $e$ for OP Tl2201 (Fig.\ \ref{8}) implies the
existence of a singular scattering mechanism, peaked at ($\pi$, 0), whose intensity grows rapidly with
decreasing carrier concentration. Moreover, the striking growth of the intensity of the scattering at ($\pi$, 0)
as we move towards the antiferromagnetic (AFM) insulating state (Fig.\ \ref{9}) points to a scattering mechanism
that is magnetic in origin, possibly due to some form of AFM spin fluctuations. Another potential candidate
though, particularly given the form of the anisotropy and its singular nature, is the collective resonance mode
seen in neutron scattering measurements on several cuprates, including Tl2201. Intriguingly, whilst the
resonance mode is resolution limited, i.e. strongly peaked at ($\pi$, 0) in Tl2201,\cite{he01} it is
significantly broadened in $k$-space in Bi2212,\cite{fong99} consistent with the conclusions derived above. It
is noted with caution however, that this mode has so far only been unambiguously resolved below $T_c$, so its
relevance to scattering in the normal state has yet to be established.

\section{Conclusions}

In summary, I have demonstrated how the seemingly anomalous dc transport properties of HTC may be accounted for
consistently within a Boltzmann transport analysis with a single $T$$^{2}$ scattering rate that varies markedly
around the basal plane. In particular, the separation of transport and Hall lifetimes, the modified Kohler's
rule and the evolution of the transport properties across the OD region of the phase diagram can all be
reproduced with a single set of parameters. In addition, the different $T$-dependence of the Hall angle observed
in the Bi-based cuprates is shown to be simply an artefact of the increased inhomogeneity of these materials;
the $T$$^2$ scattering rate continues to be the dominant source of scattering in Bi2212 but its singular nature
is smeared out by the disorder. The form of $\Gamma_{\rm{eff}}$($\phi$, $T$) is also consistent with several
features of the single-particle spectral function identified by ARPES, including the strong basal plane
anisotropy (both in the zero-temperature and the $T$-dependent scattering terms) and the tendency towards
saturation along ($\pi$, 0). Such a wide-ranging consistency with experiment is indeed encouraging and strongly
suggests that a conventional Boltzmann transport approach might be more appropriate to describe the normal state
of HTC (at least in the intermediate and low $T$ ranges) than was previously imagined.

The fact that the parent HTC compound at half-filling is an AFM Mott insulator rather than a metal has pointed
clearly to the role of strong electron correlations in the square planar CuO$_{2}$ lattice as a key element in
the cuprate problem. The introduction of holes into a highly correlated 2D magnetic background is believed to
give rise to a non-FL ground state at low doping, characterized by the highly anomalous physics of the UD
cuprates, and ultimately, unconventional high temperature superconductivity. It has often been assumed moreover,
that with continued doping, HTC eventually evolve into a conventional FL as the electron correlations become
weaker and the system becomes more three-dimensional. In this paper, I have identified a form for the normal
state scattering rate that although somewhat unconventional, is still rooted in FL physics, yet can explain the
evolution of the normal state transport properties of OD HTC right up to optimum doping. Of course, I do not
expect this simple model to remain valid on the UD side of the phase diagram, but if it shown to be applicable
to the OD and OP cuprates, then the way in which this relatively simple picture breaks down, as the pseudogap
opens and electron correlations become even stronger, might yet reveal the crucial missing link between the FL
fixed point at large doping and the curious physics of HTC on the underdoped side.

\section{Acknowledgements}

The author would like to acknowledge enlightening discussions with J.F. Annett, K. Behnia, A. Carrington, J.R.
Cooper, I.R. Fisher, B.L. Gyorffy, J.W. Loram, A.P. Mackenzie, N. Nagaosa, K.G. Sandemann, A.J. Schofield, H.
Takagi, J.L. Tallon and J.A. Wilson. The author also would like to acknowledge the hospitality of the University
of Tokyo where many of these ideas were first formulated.

\section{Appendix: Transport Coefficients for a quasi-2D Fermi surface}

The current response $\textit{\textbf{J}}_{i}$ to an applied electric $\textit{\textbf{E}}$ is given by

\[
\textit{\textbf{J}}_i=\frac{1}{4\pi^3}\int e \textit{\textbf{v}}_i g_k \textrm{d}^3 k = \sigma_{ij}
\textit{\textbf{E}}_j
\]

where $g_{k}$ is the displacement from the equilibrium distribution($g_{k}$ = $f$ - $f_{0}$) due to the applied
fields, $\sigma_{ij}$ is the conductivity tensor and $v_{i}$ is the Fermi velocity in the $i$ direction. The
total displacement can be summarized in the Bloch-Boltzmann transport equation, which under the relaxation time
approximation, is given by

\[
e \textit{\textbf{E}}\cdot \textit{\textbf{v}}_k \frac{\partial f_0}{\partial\varepsilon} +
\frac{e}{\hbar}[\textit{\textbf{v}}_k \times \textit{\textbf{B}}] \frac{\partial g_k}{\partial k} = - g_k \Gamma
\]

The various components of $\sigma_{ij}^{(n)}$ can be obtained through the Jones-Zener expansion

\[
g_k^{(n)} = (- \frac{e}{\hbar\Gamma}[\textit{\textbf{v}}_k \times \textit{\textbf{B}}] \frac{\partial}{\partial
k})^n (\frac{e \textit{\textbf{E}}\cdot \textit{\textbf{v}}_k}{\Gamma}(-\frac{\partial f_0}{\partial
\varepsilon}))
\]

to arrive at:

\[
\sigma_{ij}^{(n)} = \frac{1}{4\pi^3}\int e v_i ((-\frac{e}{\hbar\Gamma}[\textit{\textbf{v}}_k \times
\textit{\textbf{B}}]\frac{\partial}{\partial k})^n \frac{e v_j}{\Gamma}(-\frac{\partial f_0}{\partial
\varepsilon}) \textrm{d}^3 k
\]

To calculate, for instance, the in-plane Hall conductivity $\sigma_{xy}^{(1)}$, we simply apply $B_z$ and $E_y$
and calculate the response $J_x$ using the Jones-Zener term $g_k^{(1)}$.

For a quasi-2D Fermi surface, there are various techniques one can adopt to simplify the calculation. Firstly,
the element d$^3$$k$ is the product of an area parallel to the Fermi surface d$S$ and a radial (in-plane)
component d$k_r$ which is related to an incremental energy d$\varepsilon$ by

\[
\textrm{d}\varepsilon = \hbar v_F.k_F = \hbar v_F \cos \gamma \textrm{d}k_r
\]

Here $\gamma$ is the angle between $v_F$ and d$k_r$ (within the plane). Since $v_F$ is always perpendicular to
the Fermi surface, simple geometrical considerations give

\[
\gamma(\phi) = \tan ^{-1} [\frac{\partial}{\partial\phi}(\log k_F(\phi))]
\]

which is zero for a circular Fermi surface.

The element d$S$ can be expanded into the cylindrical elements $k_F$d$\phi$ and d$k_z$ to give

\[
\int (-\frac{\partial f_0}{\partial\varepsilon})\textrm{d}^3k = \int_0^{2\pi} \int_{-\pi/d}^{\pi/d}
\frac{k_F}{\hbar v_F\cos\gamma}
\]

for a quasi-2D Fermi liquid since (-$\partial f_0$/$\partial\varepsilon$) is a delta function (provided all
scattering is quasi-elastic). It is this integral $\int$ d$\phi$ that will contain all the in-plane anisotropy
of the essential parameters in our calculation. With $B
\parallel c$, the cross-product [$\textit{\textbf{v}}_k \times \textit{\textbf{B}}$] $\partial$/$\partial k$ becomes $Bv_F
\partial$/$\partial k_{\parallel}$ where

\[
\frac{\partial}{\partial k_{\parallel}} = \frac{\cos\gamma}{k_F} \frac{\partial}{\partial\phi}
\]

whilst for $B \parallel ab$, making an angle $\varphi$ with the $a$-axis

\[
[\textit{\textbf{v}}_k \times \textit{\textbf{B}}]\frac{\partial}{\partial k} = v_z[\hat{\textit{\textbf{z}}}
\times \textit{\textbf{B}}]\frac{\partial}{\partial k_{xy}} + [\textit{\textbf{v}}_{xy} \times
\textit{\textbf{B}}]\frac{\partial}{\partial k_z}
\]

i.e.

\[
[\textit{\textbf{v}}_k \times \textit{\textbf{B}}]\frac{\partial}{\partial k} \approx Bv_F \sin(\phi - \gamma -
\varphi)\frac{\partial}{\partial k_z}
\]

Finally, it is noted that the tetragonal symmetry of Tl2201 allows the limits of integration to be reduced to 0
and $\pi$/2 by including an extra factor of 4.

Employing these simple techniques, one can now derive the individual terms of the in-plane and $c$-axis
conductivity tensors. In this paper however, only the in-plane magneto-transport properties are considered and
their formulae are listed below in the isotropic case. For the anisotropic case, simply replace $v_F$, $k_F$ and
$\Gamma$ by $v_F$($\phi$), $k_F$($\phi$) and $\Gamma$($\phi$), or more appropriately for the present model, by
$\Gamma_{\rm{eff}}$($\phi$, $T$).

\[
\sigma_{xx}^{(0)} = \frac{e^2}{4\pi^3\hbar}(\frac{2\pi}{d}) 4 \int_0^{\pi/2} \frac{k_Fv_F\cos^2(\phi -
\gamma)}{\Gamma\cos\gamma} \textrm{d}\phi
\]

\[
\sigma_{xy}^{(1)} = \frac{-e^3B}{4\pi^3\hbar^2}(\frac{2\pi}{d}) 4 \int_0^{\pi/2} \frac{v_F}{\Gamma} \cos(\phi -
\gamma) \frac{\partial}{\partial\phi} (\frac{v_F}{\Gamma}\sin(\phi - \gamma))\textrm{d}\phi
\]

\[
\sigma_{xx}^{(2)} = \frac{e^4B^2}{4\pi^3\hbar^3}(\frac{2\pi}{d}) 4 \int_0^{\pi/2} \frac{v_F}{\Gamma} \cos(\phi -
\gamma)
\]
\[
\frac{\partial}{\partial\phi} \{\frac{v_F\cos\gamma}{\Gamma k_F}\frac{\partial}{\partial\phi}
(\frac{v_F}{\Gamma} \cos(\phi - \gamma))\}\textrm{d}\phi
\]

The measurable quantities $\rho_{ab}$, tan$\Theta_H$ and $\Delta\rho_{ab}$/$\rho_{ab}$ are obtained by inversion
of the conductivity tensor:

\[
\rho_{ab} = \frac{1}{\sigma_{xx}^{(0)}}
\]

\[
\tan\Theta_H = \frac{\rho_{xy}}{R_HB} \approx \frac{\sigma_{xy}^{(1)}}{\sigma_{xx}^{(0)}}
\]

\[
\frac{\Delta\rho_{ab}}{\rho_{ab}} = - \frac{\sigma_{xx}^{(2)}}{\sigma_{xx}^{(0)}} -
(\frac{\sigma_{xy}^{(1)}}{\sigma_{xx}^{(0)}})^2
\]


\begin{references}
\bibitem{bi2212data} The $\rho_{ab}$($T$) data used were the $\delta$ = 0.255 data in Fig. 2 of
ref.\cite{watanabe97}. The cot$\Theta_{H}$($T$) data were the $p$ = 0.16 sample shown in Fig. 3 of
ref.\cite{konstant00}.
\end{references}
\end{document}